\documentclass[showkeys,amsmath,amssymb,aps,preprint,groupedaddress,superscriptaddress]{revtex4-1}
\usepackage[latin1]{inputenc}

\begin{document}
\newcommand{\scalarriemann}[1]{\stackrel{#1}{R}}
\newcommand{\einstein}[2]{\stackrel{#1}{G}^{\mathrm{\raisebox{-0.1cm}{$_{#2}$}}}}
\newcommand{\connection}[2]{\stackrel{#1}{\Gamma}^{\mathrm{\raisebox{-0.1cm}{$_{#2}$}}}}
\newcommand{\connectiont}[2]{\stackrel{#1}{\omega}^{\mathrm{\raisebox{-0.1cm}{$_{#2}$}}}}

\title{Equivalent teleparallel theories in diagonalizable spacetimes: Comment on ``Metric-affine approach to teleparallel gravity''}
\author{J. B. Formiga}
\email{jansen.formiga@uespi.br}   
 \affiliation{Centro de Ciências da Natureza, Universidade Estadual do Piauí, C. Postal 381, 64002-150 Teresina, Piauí, Brazil}

\date{\today}

\begin{abstract}
It is well known that the teleparallel equivalent of general relativity yields the same vacuum solutions as general relativity does, which ensures that this particular teleparallel model is in good agreement with experiments. A less known result concerns the existence of a wider class of teleparallel models which also admits these solutions when the spacetime is diagonalizable by means of a coordinate change. However, it is stated in Ref. [Phys. Rev. D 67, 044016 (2003).] that the teleparallel equivalent of general relativity is the only teleparallel model which admits black holes. To show that this statement is not true, I prove the existence of this wider class by taking an approach different from that of Ref. [Phys. Rev. D 19, 3524 (1979)].
\end{abstract}


\maketitle

\section{Introduction}

The most attractive teleparallel model is known as teleparallel equivalent of general relativity (TEGR). This model is well known for being formally equivalent to general relativity (GR), at least in the absence of matter fields. This equivalence means that both the TEGR and GR theories have the same solutions and, therefore, they are on the same experimental footing. Nonetheless, the experiments carried out so far  do not test the validity of all these solutions. In fact, they test only the weak field approximation of these solutions, with the Schwarzschild solution playing the main role. Because of this, there is no  reason to have a teleparallel model that is equivalent to GR. It is only convenient that these models possess solutions such as the Schwarzschild and Robertson-Walker ones. Therefore, a class of models wider than the TEGR which possesses the same solution as GR does for a diagonalizable spacetime would enlarge significantly the number of models that are likely to agree with experiments.

This wider class of teleparallel models has already been proved to exist \cite{PhysRevD.19.3524}. It is intriguing to note, however, that the authors in Ref. \cite{PhysRevD.19.3524} do not emphasize this result: they do not mention it in either the abstract or the  introduction of their paper. Perhaps, that is the reason why this result is not widely known. As an example,  in Ref. \cite{PhysRevD.67.044016}, it is stated that the TEGR is the only teleparallel model which admits black holes (meaning Schwarzschild solution, too). The same statement is also present in Ref. \cite{aldrovandi2012teleparallel}. This statement not only contradicts the existence of the mentioned class, but also the results in Ref. \cite{PhysRevD.67.044016}. To clarify this point and emphasize the meaning of this wider class, I show the result obtained in Ref. \cite{PhysRevD.19.3524} by analyzing the equivalence between two teleparallel Lagrangians.

Let us see the notation used here. The holonomic and anholonomic indices are denoted by Greek and Latin letters, respectively. These indices run over $0$--$3$ and, when numbered, the anholonomic indices appear between round brackets. As an example, we have $T^a=(T^{(0)},T^{(1)},T^{(2)},T^{(3)})$. The tetrad fields are represented by $e_a$ (frame) and $e^a$ (coframe), whose components in the coordinate basis are denoted by $e_a^{\  \mu}$ and $e^a_{\  \mu}$, respectively.  As usual, the coordinate basis is denoted by $\partial_{\mu}$, which stands for $\partial/\partial x^{\mu}$ with $x^{\mu}$ being a certain coordinate system. The components of the metric tensor in the tetrad basis are $\eta_{ab}=diag(+1,-1,-1,-1)$, while the ones in the coordinate basis are denoted by $g_{\mu \nu}$. I use square brackets around indices to represent the antisymmetric part of a tensor.

My conventions are as follows. Let $M$ be a manifold endowed with a metric $g$ and a linear connection $\nabla$. Let $V,U,W$ be vectors belonging to the tangent bundle of $M$. The torsion tensor $T(V,U)$ is defined by
\begin{equation}
T(V,U)\equiv \nabla_V U-\nabla_U V-[V,U], \label{14012013a}
\end{equation}
and its components are denoted by  $T^a_{\ bc}\equiv <e^a, T(e_b,e_c)>$, where in this definition I am using the standard convention $<df,\partial_{\mu}>=\partial_{\mu} f$ with  $f$ being an arbitrary function.

\section{The misinterpretation in Ref. \cite{PhysRevD.67.044016}}

In teleparallelism there exists a special frame $e_a$ in which the affine connection coefficients vanish. This special frame and its dual basis $e^a$ will be used in all calculations performed in this paper. 

In the frame $e_a$, the torsion components  take the form:
\begin{equation}
T^a_{\  bc}=2e_b^{\  \mu}e_c^{\  \nu} e^a_{ \ [\nu,\, \mu]}, \label{2852013c}
\end{equation}
where the comma stands for partial derivative.

A general teleparallel model, sometimes also called ``new general relativity'' (NGR), can be described by the Lagrangian
\begin{equation}
T\equiv a_1T^{abc}T_{abc}+a_2T^{abc}T_{bac}+a_3T^aT_a, \label{2852013a}
\end{equation}
where $T_a\equiv T^b_{\  ba}$. 

The TEGR is the particular case where $a_1=-1/4$, $a_2=-1/2$, and $a_3=1$. So, the Lagrangian (\ref{2852013a}) becomes
\begin{equation}
\mathring{T}\equiv -\frac{1}{4}T^{abc}T_{abc}-\frac{1}{2}T^{abc}T_{bac}+T^aT_a. \label{2852013h}
\end{equation}

Let us now see the misinterpretation in Ref. \cite{PhysRevD.67.044016}. To compare the parameters ``$a$s'' defined here with those in Ref. \cite{PhysRevD.67.044016}, one may use the relations:
\begin{eqnarray}
a_1=(2\tilde{a}_1+\tilde{a}_3)/6, \quad  a_2=-(\tilde{a}_3-\tilde{a}_1)/3,
\nonumber \\
 a_3=(\tilde{a}_2-\tilde{a}_1)/3, \label{662013e}
\end{eqnarray}
where I have denoted the parameters used in Ref. \cite{PhysRevD.67.044016} with a tilde to differentiate them from the ones defined here. In this reference, the NGR is defined as being the model with the values (see section V there):
\begin{equation}
\tilde{a}_1=-1,\quad \tilde{a}_2=2, \quad \tilde{a}_3 \neq 1/2, \label{662013g}
\end{equation}
which, from the relations (\ref{662013e}), one can easily verify that it corresponds to the choice 
\begin{equation}
a_3=1,\quad a_2=-1-2a_1,\quad  a_1 \neq -1/4. \label{662013h}
\end{equation}
It is shown in this reference that all models with $\tilde{a}_1=-1$,  and $\tilde{a}_2=2$, regardless of the value of $\tilde{a}_3$, admits the Schwarzschild solution. This clearly includes what they have defined as the NGR. However,  at the end of section XII, the authors say that the TEGR is the only model which admits the Schwarzschild solution, which is a misinterpretation of their own result.

\section{The case of a diagonalizable spacetime}

The model based on the values (\ref{662013g}) not only admits the Schwarzschild solution, but also has the same vacuum solutions as GR does whenever the spacetime is diagonalizable. This was shown in Ref. \cite{PhysRevD.19.3524} by using the field equations. Here, I will prove this result by comparing the Lagrangians (\ref{2852013a}) and (\ref{2852013h}) when the spacetime is diagonalizable. 

Let us begin by proving that not all spacetimes allow for a diagonal tetrad field written in terms of a certain coordinate basis. Let $\{f_0,f_1,f_2,f_3\}$ be a set of functions. If we are able to write a tetrad field as $e^{(\mu)}=f_{\bar{\mu}}dx^{\mu}$, where $x^{\mu}$ is a particular coordinate system and the bar over the index means ``this one must not be contracted'', then the metric in these coordinates is diagonal, that is, $g=\eta_{ab}e^a\otimes e^b=\eta_{(\mu)(\nu)}f_{\bar{\mu}}f_{\bar{\nu}}dx^{\mu}\otimes dx^{\nu}$. This means that if no coordinate system diagonalizes the metric, then no diagonal tetrad can exist. In turn, it is well known that not all spacetimes have a diagonalizable metric (Kerr spacetime is an example), and hence it is not always possible to find a diagonal tetrad.

Since we can always find a diagonal tetrad for a diagonal metric, and vice versa, we can take both of them as being diagonal whenever the metric is diagonalizable by assuming that $g$ is diagonal in the coordinate system $x^{\mu}$. In this case, we can write 
\begin{eqnarray}
e^{(\alpha )}_{\ \ \mu}=f_{\bar{\mu}} \delta^{\alpha}_{\mu},\quad e_{(\alpha )}^{\ \ \mu}=f_{\bar{\mu}}^{-1} \delta^{\mu}_{\alpha},\quad g_{\mu \nu}=f_{\bar{\mu}}^2 \eta_{(\mu) (\nu)},
\nonumber \\
 g^{\mu \nu}=f_{\bar{\mu}}^{-2} \eta^{(\mu) (\nu)}.    \label{16052013b}
\end{eqnarray}
Note that when we have an index in a Kronecker delta, we can exchange its counterparts with bars for the other index in the delta. As an example, we have $B_{\bar{\mu}}C_{\bar{\mu}}\delta^{\alpha}_{\ \mu}=B_{\bar{\alpha}}C_{\bar{\alpha}}\delta^{\alpha}_{\ \mu}$. This is important when contracting indices, as in $B^{\mu}C_{\bar{\mu}}\delta^{\alpha}_{\ \mu}=B^{\alpha}C_{\bar{\alpha}}$ for example. Note, however, that $B^{\mu}g_{\mu \nu} C_{\bar{\mu}}$ is not the same as $ B_{\nu}C_{\bar{\nu}}$.

From Eq. (\ref{2852013c}), it is straightforward to verify that
\begin{equation}
T^{abc}T_{abc}=4 g^{\mu \alpha}g^{\beta \nu} e^a_{ \ [\nu,\mu]} e_{a [\beta,\alpha]} \label{2852013e}
\end{equation}
and 
\begin{equation}
T^{abc}T_{bac}=4 e_a^{ \ \alpha} e_b^{ \ \mu}  g^{\beta \nu} e^a_{ \ [\nu,\mu]} e^b_{ \ [\beta,\alpha]}. \label{2852013f}
\end{equation}
The substitution of the relations (\ref{16052013b}) into the expressions (\ref{2852013e}) and (\ref{2852013f}) lead to
\begin{equation}
T^{abc}T_{abc}=2\sum_{j=0}^{3}(f_{\bar{\mu}_j} f_j)^{-2} f_{j, \mu_j} f_{j, \alpha_j}\eta^{(\mu_j) (\alpha_j)} \label{16052013c}
\end{equation}
and
\begin{equation}
T^{abc}T_{bac}=\sum_{j=0}^{3}(f_{\bar{\mu}_j} f_j)^{-2} f_{j, \mu_j} f_{j, \alpha_j}\eta^{(\mu_j) (\alpha_j)}, \label{16052013d}
\end{equation}
where the indices with subscript run over all possible numbers except the one in the subscript, for example $\mu_1=0,2,3$. 

From Eqs. (\ref{16052013c}) and (\ref{16052013d}), we see that $T^{abc}T_{abc}=2T^{abc}T_{bac}$. By using this result in the Lagrangians (\ref{2852013a}) and (\ref{2852013h}) and assuming that $T$ equals $\mathring{T}$, we obtain
\begin{equation}
a_3 T^a T_a+(2a_1+a_2)T^{abc}T_{bac}=T^a T_a-T^{abc}T_{bac}, \label{16052013f}
\end{equation}
which is clearly satisfied for 
\begin{equation}
a_3=1, \quad 2a_1+a_2=-1. \label{1962013a}
\end{equation} 
These values include the cases (\ref{2852013h}) and (\ref{662013h}). 

For the values (\ref{1962013a}) and the spacetime (\ref{16052013b}), the Lagrangians (\ref{2852013a}) and (\ref{2852013h}) are equivalent to each other, which means that the models based on them posses the same solutions. Moreover, since both $T^{abc}T_{abc}$ and $T^{abc}T_{bac}$ are scalars, once the equation $T^{abc}T_{abc}=2T^{abc}T_{bac}$ is satisfied, it must hold in any frame and any coordinate system, including those in which the metric and tetrad fields are not diagonal. As a consequence, the Lagrangian $T$ with the values (\ref{1962013a}) yields the same field equations as GR does for any diagonalizable metric.

The existence of a coordinate change which diagonalizes the metric is not a necessary condition for the equation $T^{abc}T_{abc}=2T^{abc}T_{bac}$ to hold. As an example of a non-diagonalizable metric which satisfies this equation, we have the generic plane-wave metric \cite{0264-9381-9-7-005} 
\begin{equation}
ds^2=2du\left[dv+H(u,x,y)du\right]-dx^2-dy^2,
\end{equation}
where $H$ is an arbitrary function. It is easy to check that, for this spacetime, both (\ref{2852013e}) and (\ref{2852013f}) vanish. Therefore, the plane-wave solutions of GR are the same as that of $T$ when we take the values (\ref{1962013a}).

\section{Final remarks}

In Ref. \cite{PhysRevD.19.3524} the authors show the previous result for a diagonal metric from a completely different approach (see section VII there). They obtain the field equations in terms of the irreducible decomposition of torsion and verify that, when the spacetime is diagonal, the axial-vector part of the torsion vanishes. As a result, the field equations reduces to those of GR. It is worth mentioning that the result presented in the previous section was obtained without the knowledge of this reference, which helps to ensure its validity.
 
Perhaps, the most important feature of this result is that it enlarges the number of teleparallel models which are likely to agree with experimental tests, since the most well-established spacetimes are diagonalizable: the Schwarzschild and  Robertson-Walker ones.  It is likely that only in experiments with strong gravity one is able to detect the gravitational effects that prevent the spacetime from being diagonalizable. Even in this case, one will not be able to distinguish these models if the experiments deal with the detection of plane gravitational waves.


\begin{thebibliography}{4}
\expandafter\ifx\csname natexlab\endcsname\relax\def\natexlab#1{#1}\fi
\expandafter\ifx\csname bibnamefont\endcsname\relax
  \def\bibnamefont#1{#1}\fi
\expandafter\ifx\csname bibfnamefont\endcsname\relax
  \def\bibfnamefont#1{#1}\fi
\expandafter\ifx\csname citenamefont\endcsname\relax
  \def\citenamefont#1{#1}\fi
\expandafter\ifx\csname url\endcsname\relax
  \def\url#1{\texttt{#1}}\fi
\expandafter\ifx\csname urlprefix\endcsname\relax\def\urlprefix{URL }\fi
\providecommand{\bibinfo}[2]{#2}
\providecommand{\eprint}[2][]{\url{#2}}

\bibitem[{\citenamefont{Hayashi and Shirafuji}(1979)}]{PhysRevD.19.3524}
\bibinfo{author}{\bibfnamefont{K.}~\bibnamefont{Hayashi}} \bibnamefont{and}
  \bibinfo{author}{\bibfnamefont{T.}~\bibnamefont{Shirafuji}},
  \bibinfo{journal}{Phys. Rev. D} \textbf{\bibinfo{volume}{19}},
  \bibinfo{pages}{3524} (\bibinfo{year}{1979}),
  \urlprefix\url{http://link.aps.org/doi/10.1103/PhysRevD.19.3524}.

\bibitem[{\citenamefont{Obukhov and Pereira}(2003)}]{PhysRevD.67.044016}
\bibinfo{author}{\bibfnamefont{Y.~N.} \bibnamefont{Obukhov}} \bibnamefont{and}
  \bibinfo{author}{\bibfnamefont{J.~G.} \bibnamefont{Pereira}},
  \bibinfo{journal}{Phys. Rev. D} \textbf{\bibinfo{volume}{67}},
  \bibinfo{pages}{044016} (\bibinfo{year}{2003}),
  \urlprefix\url{http://link.aps.org/doi/10.1103/PhysRevD.67.044016}.

\bibitem[{\citenamefont{Aldrovandi and
  Pereira}(2012)}]{aldrovandi2012teleparallel}
\bibinfo{author}{\bibfnamefont{R.}~\bibnamefont{Aldrovandi}} \bibnamefont{and}
  \bibinfo{author}{\bibfnamefont{J.}~\bibnamefont{Pereira}},
  \emph{\bibinfo{title}{Teleparallel Gravity: An Introduction}}
  (\bibinfo{publisher}{Springer, London}, \bibinfo{year}{2012}), ISBN
  \bibinfo{isbn}{9789400751422}.

\bibitem[{\citenamefont{Tod}(1992)}]{0264-9381-9-7-005}
\bibinfo{author}{\bibfnamefont{K.~P.} \bibnamefont{Tod}},
  \bibinfo{journal}{Classical and Quantum Gravity}
  \textbf{\bibinfo{volume}{9}}, \bibinfo{pages}{1693} (\bibinfo{year}{1992}),
  \urlprefix\url{http://stacks.iop.org/0264-9381/9/i=7/a=005}.

\end{thebibliography}

\end{document}